\documentclass{PoS}
\usepackage{epsfig,amsfonts,amsthm}
\usepackage[normalem]{ulem}
\usepackage{amsmath,amssymb}
\usepackage{array}
\usepackage{amsmath}
\usepackage{amsfonts}
\usepackage{amssymb}
\usepackage{subfig}
\usepackage{wrapfig}
\usepackage{graphicx}
\newcommand{\be}{\begin{equation}}
\newcommand{\ee}{\end{equation}}
\newcommand{\bea}{\begin{eqnarray}}
\newcommand{\eea}{\end{eqnarray}}

\renewcommand{\Re}{\mathrm{Re }}
\renewcommand{\Im}{\mathrm{Im }}
\newcommand{\doublet}[2]{ \left( \begin{array}{c}#1 \\ #2 \end{array}\right) }

\usepackage{color}

\def\lsim{\mathrel{\rlap{\lower4pt\hbox{\hskip1pt$\sim$}}
    \raise1pt\hbox{$<$}}}         %less than or approx. symbol
\def\gsim{\mathrel{\rlap{\lower4pt\hbox{\hskip1pt$\sim$}}
    \raise1pt\hbox{$>$}}}         %greater than or approx. symbol

\def\beq{\begin{equation}}
\def\eeq{\end{equation}}
\def\bea{\begin{eqnarray}}
\def\eea{\end{eqnarray}}

\def\<{\left\langle}
\def\>{\right\rangle}

\usepackage[normalem]{ulem}

\def\lsim{\mathrel{\rlap{\lower4pt\hbox{\hskip1pt$\sim$}}
    \raise1pt\hbox{$<$}}}         %less than or approx. symbol
\def\gsim{\mathrel{\rlap{\lower4pt\hbox{\hskip1pt$\sim$}}
    \raise1pt\hbox{$>$}}}         %greater than or approx. symbol

\def\beq{\begin{equation}}
\def\eeq{\end{equation}}
\def\bea{\begin{eqnarray}}
\def\eea{\end{eqnarray}}

\def\<{\left\langle}
\def\>{\right\rangle}

\newcommand{\bt}{\begin{tabular}}
\newcommand{\et}{\end{tabular}}

\title{CP violation and BSM Higgs bosons}

\ShortTitle{CPV and BSM Higgses}

\author{\speaker{Venus Keus}\\
        Department of Physics and Helsinki Institute of Physics, \\
        Gustaf Hallstromin katu 2, FIN-00014 University of Helsinki, Finland\\
        E-mail: \email{venus.keus@helsinki.fi}}

\abstract{We study extensions of the Standard Model (SM) in which copies of the SM scalar $SU(2)$ doublet are added to the Higgs sector. These scalar doublets either acquire a Vacuum Expectation Value (VEV) and hence are \textit{active} or do not develop a VEV and are \textit{inert}. We consider CP-violation (CPV) in both the active and inert sector. As an example of a model with CPV in the active sector, we present a Type-I 2-Higgs-Doublet Model (2HDM) with two active doublets and show Large Hadron Collider (LHC) signals of such a scenario. The amount of CPV in this case is very limited due to constraints coming from Electric Dipole Moment experiments. Moreover, 2HDMs with only active doublets do not provide a Dark Matter (DM) candidate. As a result, we turn to 3-Higgs-Doublet Models (3HDMs) where unbounded CPV and viable DM candidates could be introduced simultaneously in the inert sector. We investigate DM phenomenology of such models.}

\FullConference{Prospects for Charged Higgs Discovery at Colliders\\
		3-6 October 2016\\
		Uppsala, Sweden}

\begin{document}

\section{Introduction}
The great success of the Standard Model (SM) was achieved in 2012 with the discovery of the long-awaited Higgs boson \cite{Aad:2012tfa, Chatrchyan:2012ufa}, the last missing particle of the SM at the large Hadron Collider (LHC), as this particle was predicted by the Electro-Weak Symmetry Breaking (EWSB) mechanism in 1964. 
Even though no significant deviation from the SM has been observed so far, it is well understood that the SM of particle physics is not the complete theory of Nature. Many astrophysical observations require a Dark Matter (DM) which is stable on cosmological time scales, cold, non-baryonic, neutral and weakly interacting. The SM does not provide any such candidate. The most well-studied such candidates are the Weakly Interacting Massive Particles (WIMPs) \cite{Jungman:1995df,Bertone:2004pz,Bergstrom:2000pn}, with masses between a few GeV and a few TeV. 
Such a WIMP candidate must be cosmologically 
stable, usually due to the conservation of a discrete symmetry, and must freeze-out to result in the 
observed relic density \cite{Ade:2015xua}:
\begin{equation}\label{relic}
\Omega_{\rm DM} h^2 = 0.1199 \pm 0.0027. 
\end{equation}
Similarly the Baryon asymmetry of the Universe (BAU) which requires the CP-symmetry to be broken, cannot be explained by the SM since the amount of CP-violation (CPV) in SM does not lead to enough matter-antimatter asymmetry to match observational data \cite{dimopoulos,Cline:2006ts}. 
%On the other hand, the value of Higgs mass in the SM is not compatible with a first order phase transition in the early Universe, which is a necessary requirement for explaining BAU. 

The simplest Beyond Standard Model (BSM) scenarios which aim to conquer these shortcomings are scalar extensions. 2-Higgs-Doublet Models (2HDMs) with one scalar $SU(2)$ doublet added to the SM could partly provide an answer. A 2HDM with two \textit{active} doublets could accommodate CPV. In section~\ref{2hdm}, we study a Type-I 2-Higgs-Doublet Model (2HDM) with CPV and present signatures of this model for future discovery at the LHC. Since both doublets are active, the amount of CPV introduced here is very limited due to bounds coming from Electric Dipole Moment (EDM) experiments. Moreover, 2HDMs with CPV fall short on providing a DM candidate. 

The Inert Doublet Model (IDM) \cite{Deshpande:1977rw} with one \textit{inert} and one \textit{active} doublet in the scalar sector, provides a viable DM candidate, the lightest neutral particle from the inert doublet. The IDM has been studied extensively in the last few years (see, e.g., \cite{Ma:2006km,Barbieri:2006dq,LopezHonorez:2006gr}) where a discrete $Z_2$ symmetry unbroken after EWSB protects the DM from decaying into SM fields. 
Since the IDM involves {\em 1} Inert Doublet plus {\em 1} active Higgs Doublet, we shall also refer to it henceforth as the I(1+1)HDM. 
The I(1+1)HDM, though very constrained, remains a viable model for a scalar DM candidate, being in agreement with current experimental constraints.
However, by construction, the I(1+1)HDM can not contain CPV; due to the presence of an exact $Z_2$ symmetry, all parameters in the potential are real. To incorporate both DM and CPV, we need to go beyond two scalar doublets. 

In section~\ref{3hdm}, we study a 3-Higgs-Doublet model (3HDM) with {\em 2} inert Higgs plus {\em 1} active Higgs doublet which provides a viable DM candidate and unbounded amount of CPV arising from the inert sector. We refer to this model as the I(2+1)HDM. Here, the active sector is identical to that of the SM and the inert sector is extended. CPV is introduced in the inert sector and the neutral inert particles now have a mixed CP quantum number.

\section{CP-violating 2-Higgs-Doublet Models}\label{2hdm}

\subsection{The scalar potential}

In the general form, the 2HDMs allow for Flavour Changing Neutral Currents (FCNCs) at the tree level which have not been observed experimentally. To forbid these FCNCs, the scalar potential could be restricted by imposing a $Z_2$ symmetry which is also extended to the fermion sector. 
Depending on this $Z_2$ charge assignment on the fermions, four independent types of 2HDMs are defined \cite{four-types,typeX}. Here, we study the Type-I model where only one Higgs doublet, $\phi_1$, couples to fermions. The $Z_2$ charges of $\phi_1$ and $\phi_2$ is set to be even and odd, respectively. 

The $Z_2$ symmetric 2HDM potential has the form:
\bea
\label{potgen}
V &=&  
\mu^2_{1} (\phi_1^\dagger \phi_1) +\mu^2_2 (\phi_2^\dagger \phi_2) -\biggl[ \mu^2_3(\phi_1^\dagger \phi_2) + h.c. \biggr] 
+ \frac{1}{2} \lambda_{1} (\phi_1^\dagger \phi_1)^2 +\frac{1}{2} \lambda_{2} (\phi_2^\dagger \phi_2)^2  
\nonumber\\
&& 
+ \lambda_{3} (\phi_1^\dagger \phi_1)(\phi_2^\dagger \phi_2)  + \lambda_{4} (\phi_1^\dagger \phi_2)(\phi_2^\dagger \phi_1)
+  \biggl[\frac{1}{2} \lambda_{5}  (\phi_1^\dagger \phi_2)^2 + h.c. \biggr]. 
\eea
with the doublets compositions defined as
\be 
\phi_1= \doublet{$\begin{scriptsize}$ \phi^+_1 $\end{scriptsize}$}{\frac{v_1+h_1^0+ia^0_1}{\sqrt{2}}} ,\quad 
\phi_2= \doublet{$\begin{scriptsize}$ \phi^+_2 $\end{scriptsize}$}{\frac{v_2+h^0_2+ia^0_2}{\sqrt{2}}} ,
\label{fields}
\ee
where $v_1$ and $v_2$ could be complex. 
We allow for the $Z_2$ soft-breaking term, $\mu_3^2$ and take the Vacuum Expectation Values (VEVs) to be real and positive and defined as $v^2\equiv v_1^2+v_2^2 = (246$ GeV$)^2$ and the ratio of the two VEVs is $\tan\beta =v_2/v_1$. 
Therefore, CPV is introduced explicitly through the only complex terms in the potential, $\mu^2_3$ and $\lambda_5$. The minimisation conditions require
\be 
\Im\mu_3^2 = \frac{v^2}{2}\Im\lambda_5 s_\beta  c_\beta.
\ee
Therefore $\Im\lambda_5$ may be regarded as the only source of CPV. 
Throughout the paper, we will be using the following
abbreviations: $s_\theta = \sin\theta$,  $c_\theta = \cos\theta$
and $t_\theta = \tan\theta$. 

We make use of the Higgs basis \cite{HiggsBasis} where the rotated doublets are represented by $\hat \phi_i$
\be 
\left(\begin{array}{c}
\hat \phi_1 \\ 
\hat \phi_2
\end{array}\right)
= \left(\begin{array}{cc}
c_\beta & s_\beta\\
-s_\beta & c_\beta
\end{array}\right)
\left(\begin{array}{c}
\phi_1 \\ 
\phi_2
\end{array}\right), 
\ee
where $ \langle \hat \phi_1 \rangle =v$ and $ \langle \hat \phi_2 \rangle = 0$.

The mass of the charged Higgs boson, $H^\pm$, is calculated to be
\be 
m^2_{H^\pm}= \frac{\Re\mu_3^2}{s_\beta c_\beta} -\frac{v^2}{2}(\lambda_4 + \Re\lambda_5). 
\ee
The masses of the three neutral CP-mixed states, $H_1$, $H_2$ and $H_3$, is calculated by diagonalising the $3 \times 3$ mass-squared matrix where we assume 
$m_{H_1}^{}\leq m_{H_2}^{}\leq m_{H_3}^{}$. The explicit form of the rotation matrix and Feynman rules are presented in \cite{Keus:2015hva}.
In the following, we identify $H_1$ as the SM-like Higgs boson, so that we take $m_{H_1}^{}=125$ GeV.

\subsection{Constraints on the model}

We have taken into account the stability of the Higgs potential, S-matrix unitarity and the $S$, $T$ and $U$ parameters \cite{Peskin-Takeuchi,STU-THDM}. 
We also take into account EDMs which limit the CPV parameter, i.e., $\Im\lambda_5$ significantly.
The $B$ physics data also provides constraints on the parameter space in 2HDMs, which are especially sensitive to $m_{H^\pm}$ and $\tan\beta$. 
In addition, we also take into account the constraint from direct searches for extra Higgs bosons at the LHC. 

\begin{figure}[h!]
\begin{center}
\includegraphics[scale=0.35]{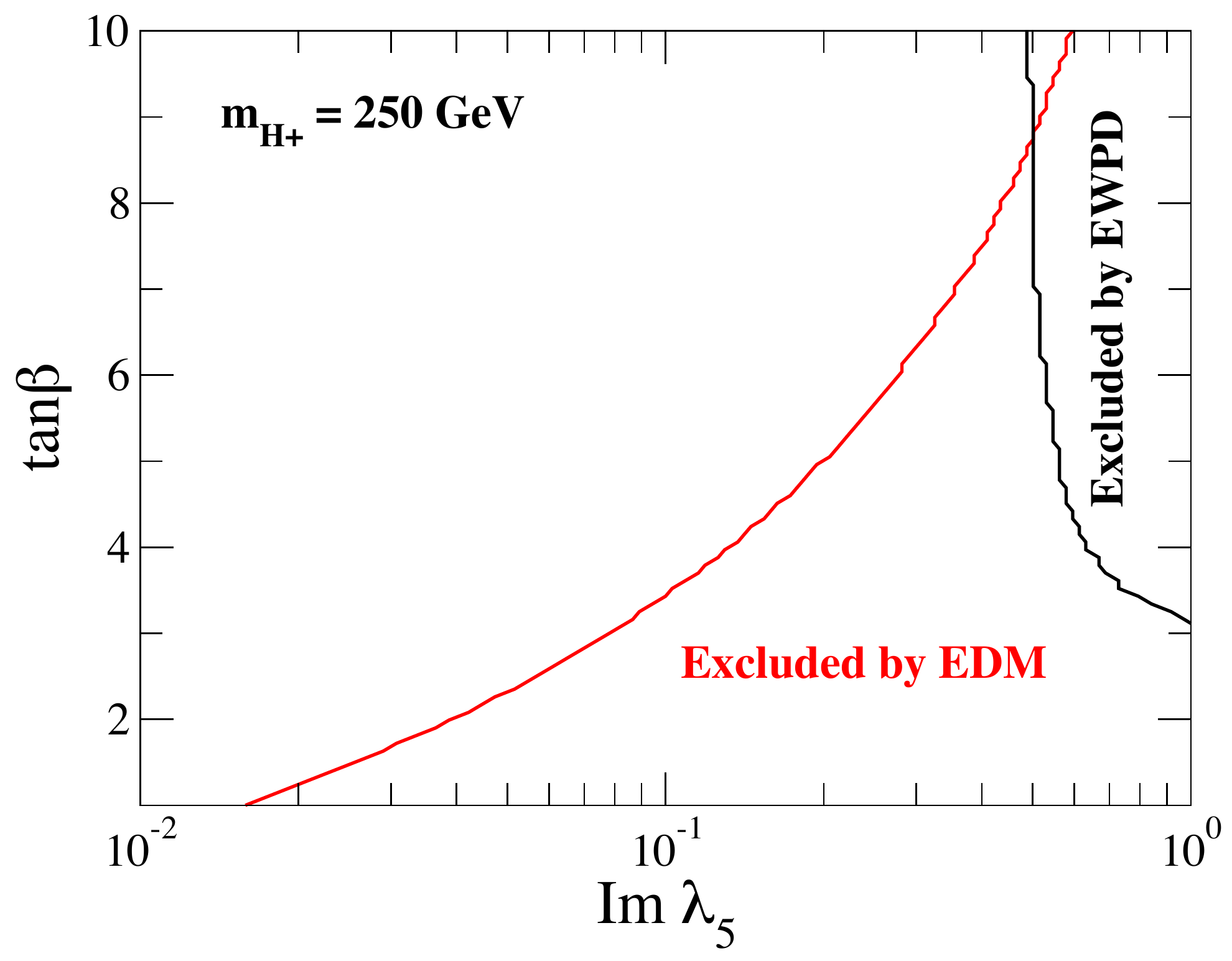}
\caption{The constrained region in the $\Im\lambda_5$-$\tan\beta$ plane is shown in the case of ${m}_{H_2}$ = 200 GeV, ${m}_{H_3} = m_{H^\pm}$ = 250 GeV and $s_{\beta-\tilde{\alpha}}=1$.}
\label{bounds1}
\end{center}
\end{figure}

Fig.~\ref{bounds1} summarizes the bounds taken into account where we show the allowed parameter regions on the $\Im\lambda_5$ and $\tan\beta$ plane. In this plot, we take ${m}_{H_2}$ = 200 GeV, ${m}_{H_3} = m_{H^\pm}$ = 250 GeV and $s_{\beta-\tilde{\alpha}}=1$ where $\beta-\tilde{\alpha}$ is the mixing angle which diagonalises the CP-even Higgs states in the Higgs basis.

\subsection{Phenomenology at the LHC}

\begin{figure}[h!]
\begin{center}
\includegraphics[scale=0.26]{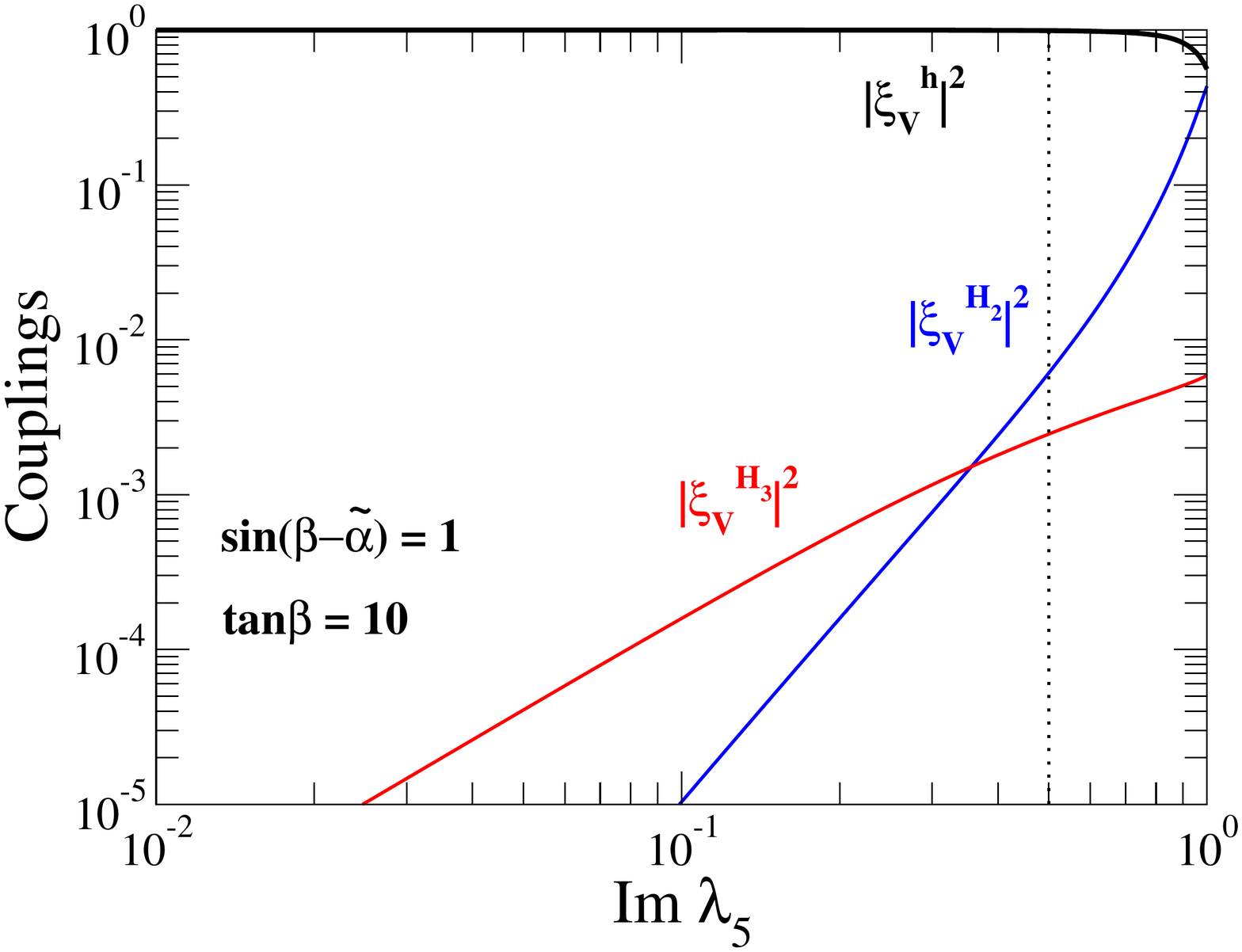}
\hspace{-1mm} 
\includegraphics[scale=0.26]{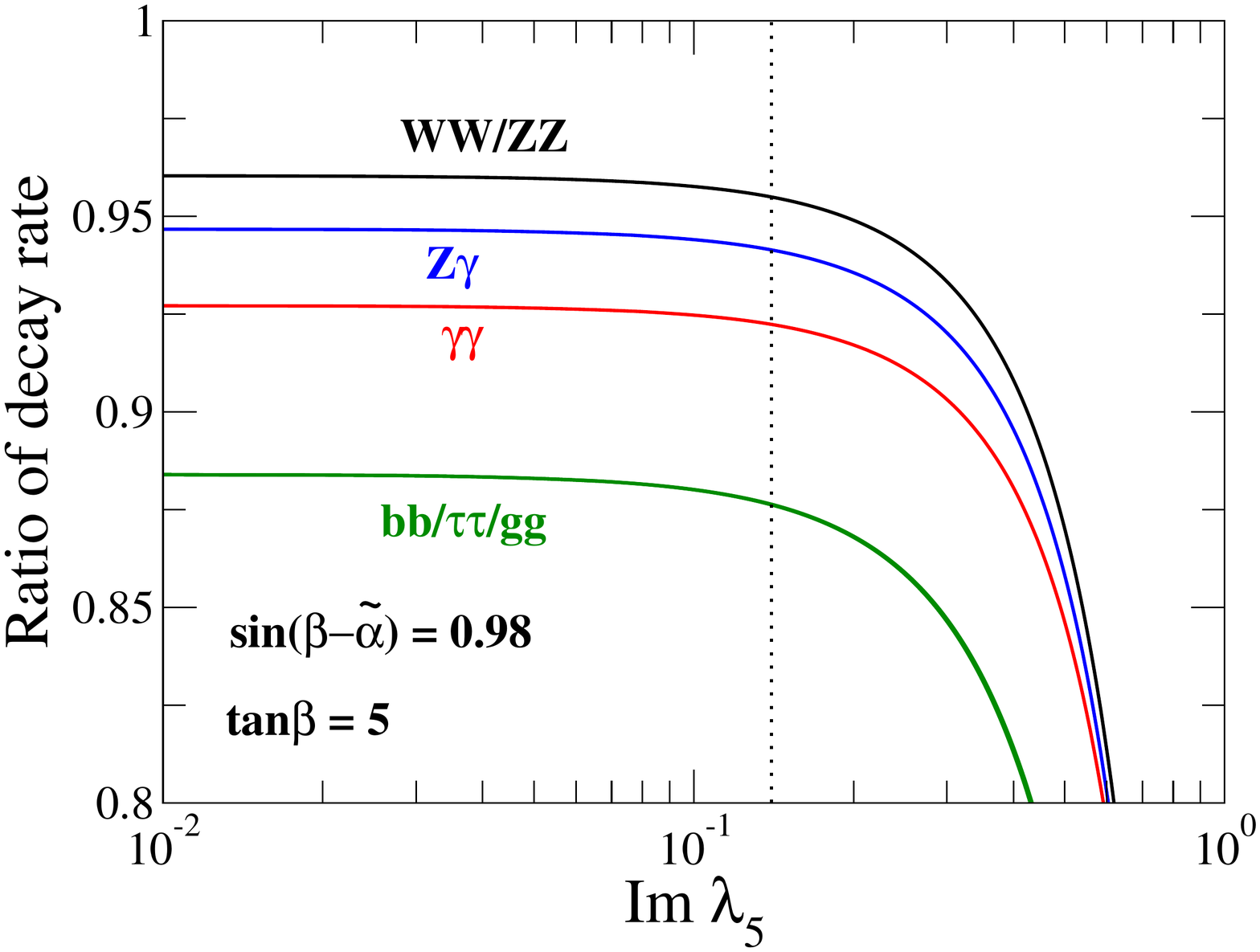} \\
\vspace{-3mm}
\includegraphics[scale=0.26]{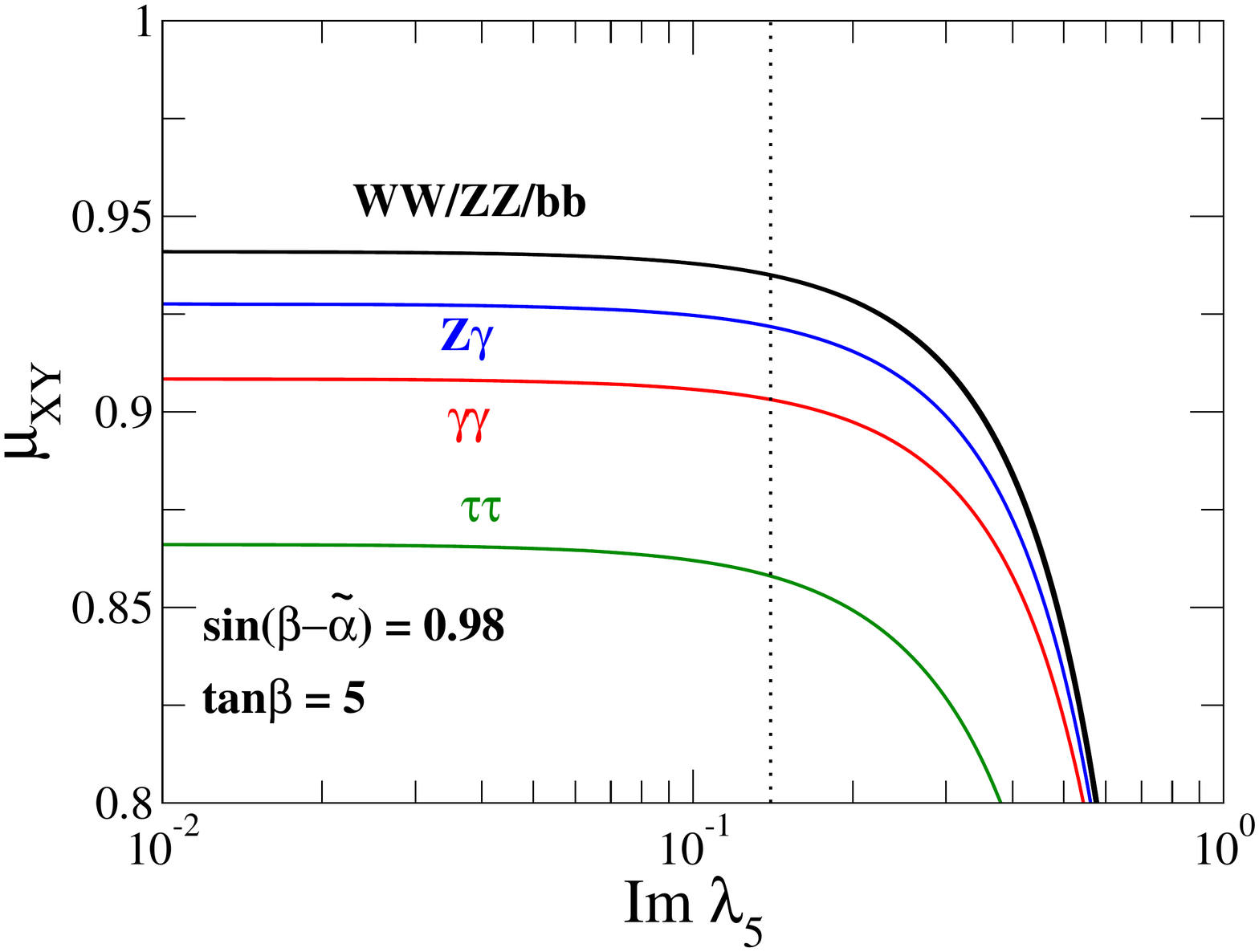}  
\caption{Signature of CPV in different regions of the parameter space.}
\label{signals}
\end{center}
\end{figure}

For our numerical results, we use the fixed input parameters ${m}_{H_2}=200$ GeV and ${m}_{H_3}=m_{H^\pm}=250$ GeV. In the top left plot in Fig. \ref{signals}, we show the gauge-gauge-scalar couplings which are described by 
$g_{hVV}^{\text{SM}}\times \xi_V^{H_i}$ ($i=1,2,3$) as a function of $\Im\lambda_5$ where $\tan\beta=10$ and $s_{\beta-\tilde{\alpha}}=1$.
The vertical dotted line shows the upper limit on $\Im\lambda_5$. 
It is evident that, over the allowed values of  $\Im\lambda_5$, deviations of the SM-like Higgs couplings to $W^+W^-$ and $ZZ$ induced by CPV 
are negligible, thereby generating no tension against LHC data. We establish the $W^+W^-$ and $ZZ$ decays of all three Higgs states of the 2HDM Type-I as a hallmark signature of CPV.

On the top right plot in Fig. \ref{signals}, we present the ratio of decay rates of the $H_1$ (identified as the SM-like Higgs boson) to those of $h_{\text{SM}}$ (the Higgs boson in the SM) 
for $\tan\beta=5$ and $ s_{\beta-\tilde{\alpha}}=0.98$.
Over the allowed $\Im\lambda_5$ region, none of BRs of the SM-like Higgs boson deviates significantly from the LHC data, except for $b\bar b, \tau^+\tau^-$ and $gg$. Therefore, this effect may be significant in order
to establish CPV. Note that this occurs in a complementary region of the parameter space in comparison to where the $W^+W^-$ and $ZZ$ signals of the neutral Higgs states can be seen.

The bottom plot in Fig. \ref{signals} shows the signal strength, $\mu_{XY}^{}$, of the SM-like Higgs boson $H_1$. CPV affects the signal strengths via the production cross
sections, partial decay widths and the total decay width. Here, only the $\tau^+\tau^-$ channel may carry some evidence of CPV for $s_{\beta-\tilde{\alpha}}=0.98$ and $\tan\beta=5$. Hence, this offers a second
channel to access CPV in the 2HDM Type-I studied here, alternative to the smoking gun signature of 
$W^+W^-$ and $ZZ$ decays.

%%%%%%%%%%%%%%%%%%%%%%%%%%%%
\section{CP-violating 3-Higgs-Doublet Models}\label{3hdm}

\subsection{The scalar potential}
\label{scalar-potential}

In this section we study a 3HDMs with {\em 2} inert Higgs plus {\em 1} active Higgs doublet which we refer to as the I(2+1)HDM. 
It has been shown \cite{Ivanov:2011ae} that a $Z_2$-symmetric 3HDM potential\footnote{Note that adding extra $Z_2$-respecting terms such as
$ 
(\phi_3^\dagger\phi_1)(\phi_2^\dagger\phi_3), 
(\phi_1^\dagger\phi_2)(\phi_3^\dagger\phi_3), 
(\phi_1^\dagger\phi_2)(\phi_1^\dagger\phi_1)$ and/or 
$(\phi_1^\dagger\phi_2)(\phi_2^\dagger\phi_2)
$
does not change the phenomenology of the model. The coefficients of these terms, therefore, have been set to zero for simplicity.}, under whose symmetry the three Higgs doublets $\phi_{1,2,3}$ transform, respectively, as 
$g_{Z_2}=  \mathrm{diag}\left(-1, -1, 1 \right)$,
is of the following form:
\bea
\label{V0-3HDM}
V_{3HDM}&=& - \mu^2_{1} (\phi_1^\dagger \phi_1) -\mu^2_2 (\phi_2^\dagger \phi_2) - \mu^2_3(\phi_3^\dagger \phi_3)+ \lambda_{11} (\phi_1^\dagger \phi_1)^2+ \lambda_{22} (\phi_2^\dagger \phi_2)^2  + \lambda_{33} (\phi_3^\dagger \phi_3)^2 \nonumber\\
&& + \lambda_{12}  (\phi_1^\dagger \phi_1)(\phi_2^\dagger \phi_2)
 + \lambda_{23}  (\phi_2^\dagger \phi_2)(\phi_3^\dagger \phi_3) + \lambda_{31} (\phi_3^\dagger \phi_3)(\phi_1^\dagger \phi_1) \nonumber\\
&& + \lambda'_{12} (\phi_1^\dagger \phi_2)(\phi_2^\dagger \phi_1) 
 + \lambda'_{23} (\phi_2^\dagger \phi_3)(\phi_3^\dagger \phi_2) + \lambda'_{31} (\phi_3^\dagger \phi_1)(\phi_1^\dagger \phi_3),  \nonumber\\
&& -\mu^2_{12}(\phi_1^\dagger\phi_2)+  \lambda_{1}(\phi_1^\dagger\phi_2)^2 + \lambda_2(\phi_2^\dagger\phi_3)^2 + \lambda_3(\phi_3^\dagger\phi_1)^2  + h.c. \nonumber
\eea
where CPV is introduced explicitly through complex parameters of the potential.

The doublets are defined as
\be 
\phi_1= \doublet{$\begin{scriptsize}$ H^+_1 $\end{scriptsize}$}{\frac{H^0_1+iA^0_1}{\sqrt{2}}},\quad 
\phi_2= \doublet{$\begin{scriptsize}$ H^+_2 $\end{scriptsize}$}{\frac{H^0_2+iA^0_2}{\sqrt{2}}}, \quad 
\phi_3= \doublet{$\begin{scriptsize}$ G^+ $\end{scriptsize}$}{\frac{v+h+iG^0}{\sqrt{2}}}, 
\label{explicit-fields}
\ee
where $\phi_1$ and $\phi_2$ are the two \textit{inert} doublets (odd under the $Z_2$) and $\phi_3$ is the one \textit{active} doublet (even under the $Z_2$) which plays the role of the SM-Higgs doublet. The symmetry of the potential is therefore respected by the vacuum alignment.

To make sure that the entire Lagrangian is $Z_2$ symmetric, we assign an even $Z_2$ parity to all SM particles, identical to the $Z_2$ parity of the only doublet that couples to them, i.e., the active doublet $\phi_3$ \cite{Ivanov:2012hc}. With this parity assignment FCNCs are avoided as the extra doublets are forbidden to couple to fermions by $Z_2$ conservation.

We take $S_1$ to be the lightest neutral field from the inert doublets which now have a mixed CP-charge:
\bea 
\label{masses}
&& 
S_1 =\frac{\alpha H_1^0 + \alpha H_2^0-A_1^0+A_2^0}{\sqrt{2\alpha^2+2}},\quad
S_2 =\frac{-H_1^0-H_2^0 -\alpha A_1^0+ \alpha A_2^0}{\sqrt{2\alpha^2+2}},  \\
&& 
S_3 =\frac{\beta H_1^0 -\beta H_2^0+A_1^0+A_2^0}{\sqrt{2\beta^2+2}},
\quad
S_4 =\frac{- H_1^0+ H_2^0 +\beta A_1^0 +\beta A_2^0}{\sqrt{2\beta^2+2}}, \nonumber
\eea
and hence the DM candidate
where $\alpha$ and $\beta$ are the rotation angles \cite{Cordero-Cid:2016krd}. 
Here, we study a simplified version of the I(2+1)HDM by imposing the following equalities
$
\mu^2_1 =\mu^2_2 , \lambda_3=\lambda_2 ,  \lambda_{31}=\lambda_{23} , \lambda'_{31}=\lambda'_{23} 
$
which is sometimes referred to as the ``dark democracy'' limit. By imposing the ``dark democracy'' limit, the only two parameters that remain complex are $\mu^2_{12}$ and $\lambda_2$ with $\theta_{12}$ and $\theta_2$ as their CPV phases, respectively. 
Note that the \textit{inert} sector is protected by a conserved $Z_2$ symmetry from coupling to the SM particles, therefore, the amount of CPV introduced here is not constrained by SM data, unlike what was presented in section \ref{2hdm}. 

\subsection{Constraints on parameters}\label{constraints}

We take into account constraints that the boundedness of the potential, positive-definiteness of the Hessian and $S,T,U$ parameters put on the model. Properties of all inert scalars are constrained by various experimental results. We have considered bounds from relic density observations, Gamma-ray searches, DM direct and indirect detection, the contribution of the new scalars to the $W$ and $Z$ gauge boson widths, null searches for charged scalars, invisible Higgs decays, Higgs total decay width and the $h\to \gamma \gamma$ signal strength.

\subsection{Dark Matter relic density}\label{coannihilation}

Taking all (co)annihilation processes into account, we present three benchmark scenarios, A, B and C, in the low ($m_{S_1} < m_h/2$) and medium mass region ($<m_h/2 < m_{S_1} < m_Z$) :
Scenario A where
$m_{S_1} \ll m_{S_2}, m_{S_3}, m_{S_4}, m_{S^\pm_1}, m_{S^\pm_2}$, 
scenario B where
$
m_{S_1} \sim m_{S_3} \ll m_{S_2},  m_{S_4}, m_{S^\pm_1}, m_{S^\pm_2}$, and 
scenario C where
$
m_{S_1} \sim m_{S_3} \sim m_{S_2} \sim  m_{S_4} \ll m_{S^\pm_1}, m_{S^\pm_2}$. For the numerical details of these benchmarks scenarios see \cite{Cordero-Cid:2016krd}.

In the CP-conserving version of the I(2+1)HDM (within the ``dark democracy'' limit) \cite{Keus:2014jha, Keus:2015xya}, the inert scalar-gauge couplings are fixed, and given by the rotation angles $\theta_a=\theta_h=\pi/4$. They do not depend on the mass splittings or the value of $m_{S_1}$. In the CPV case, however, these couplings depend on the rotation angles $\alpha$ and $\beta$ in Eq.~(\ref{masses}), which in turn depend on $m_{S_i}$. Higgs-inert scalar couplings are also modified with respect to the CP-conserving case which leads to different DM and LHC phenomenology of the model.

\begin{figure}[h]
\centering
\includegraphics[scale=1.1]{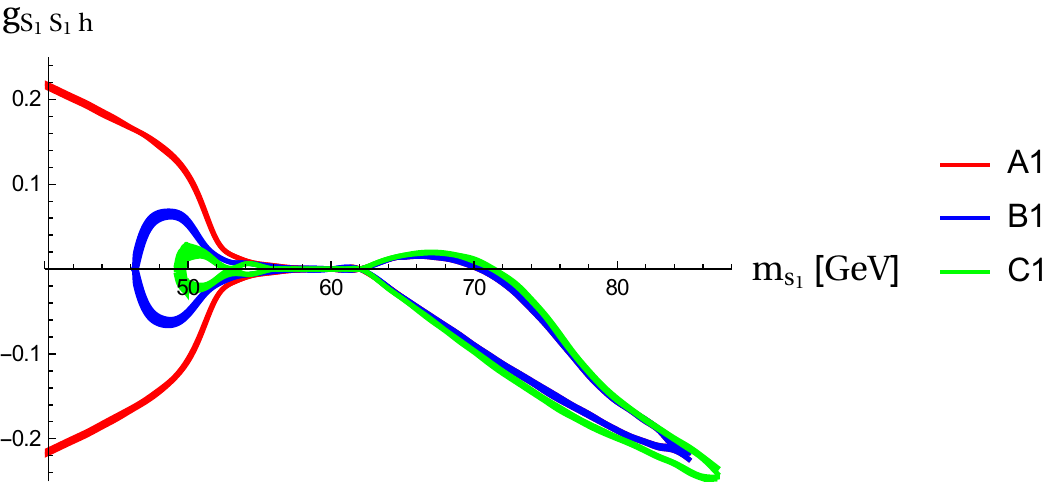}
\caption{Relic density for low DM mass region in Scenarios A, B and C. \label{relic1}}
\end{figure}

In Fig. \ref{relic1}, we show values of DM mass and Higgs-DM coupling that lead to the correct DM relic density for benchmarks A, B and C. Benchmark A with no coannihilation channels presents the standard behaviour of an $SU(2)$ DM candidate. In benchmark B for large values of $g_{S_1S_1h}$ the dominant annihilation channel is $S_1S_1 \to \bar{b}b$ and, as there are also coannihilation channels, the relic density is usually too small. For smaller couplings the dominant channel is $S_1 S_3 \to Z \to q \bar{q}$ where the relevant cross section is too large. As the mass grows, the coannihilation channel gets weaker, allowing DM to obtain the proper relic density. For masses closer to $m_h/2$ the resonance annihilation dominates, following the same pattern as in benchmark A. In benchmark C, for small values of $g_{S_1S_1h}$ the dominant coannihilation channel is $S_1 S_4 \to Z \to f \bar{f}$ (light quarks), with a small contribution from $S_2 S_3 \to Z \to f \bar{f}$. For larger couplings the process $S_1 S_1 \to h \to b \bar{b}$ strongly increases the annihilation cross section.

In all scenarios, when DM mass is close to $m_h/2$ the Higgs-resonance annihilation takes over and only very small Higgs-DM couplings lead to the correct relic density value.

In the medium DM mass, $m_h/2 < m_{S_1} < m_{W^\pm,Z}$ the crucial channel for all benchmarks is the quartic process $S_1 S_1 \to W^+ W^-$ which does not depend on the rotation angles $\alpha$ and $\beta$. For this reason, all studied benchmarks as well as the CP-conserving scenarios follow a similar behaviour. For larger values of DM mass this annihilation is stronger, and cancellation with $S_1 S_1 \to h \to W^+ W^-$ is needed to get the proper relic density value. Therefore, the plot moves towards negative values of Higgs-DM coupling. In benchmarks B and C, other channels, such as $S_1 S_4 \to q \bar{q}$ or $S_3 S_3 \to W^+ W^-$ have a small contributions, leading to small deviations from the behaviour of benchmark A.

\begin{figure}[htb]
\centering
\includegraphics[scale=1.2]{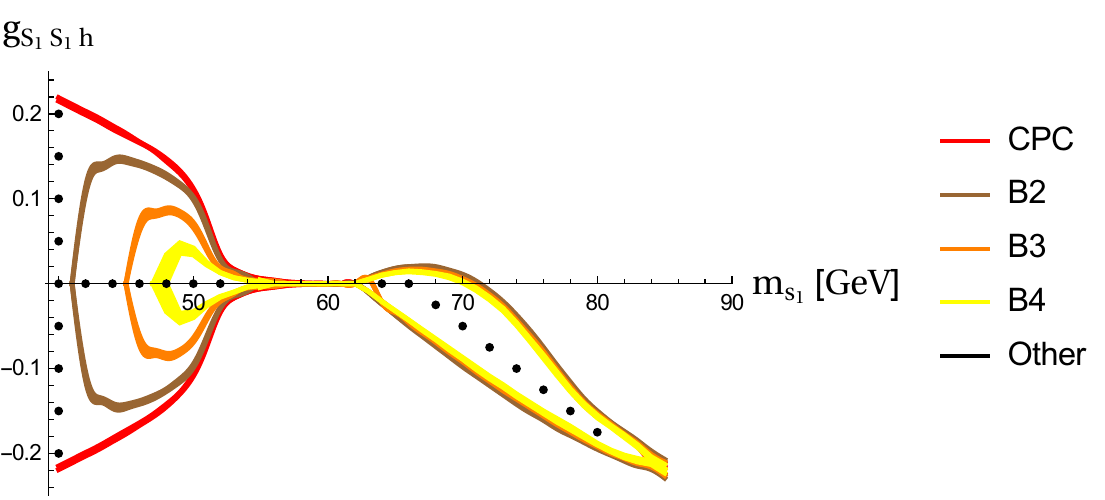}
\caption{The relic density plots for different B and C scenarios where by changing the angles $\theta_2$ and $\theta_{12}$ the whole region not accesible by the CP-conserving limit could be realised in the CPV case.\label{filling}}
\end{figure}

In Fig. \ref{relic1}, we have presented results for three sets of parameters in scenarios A, B, and C. It is clear that by changing the input set we can reach different regions of parameters space. Compare, for example, scenarios A and B, which differ only by the chosen values of $\theta_2$ and $\theta_{12}$. The performed scan shows that by varying the mass splitting and phases $\theta_2$ and $\theta_{12}$ we can actually fill the empty regions inside the plots in Fig. \ref{relic1}. In Fig. \ref{filling} results obtained for various additional sets of parameters are presented. We can fill the plot by different B and C scenarios, where the contribution from the coannihilation channels is crucial.

This is the where the affect of CPV is evident. Direct and indirect detection experiments as well as LHC limits constrain the Higgs-DM coupling severely in all Higgs-portal DM type models. However, in the CPV I(2+1)HDM, owing to the freedom in the strength of inert-gauge couplings, one can opt for very small Higgs-DM couplings while getting the proper relic density and surviving all DM detection and LHC experiments.

\section{Conclusion}\label{conclusion}

We have looked into scalar extensions of the SM with CPV. We have presented LHC signatures of a Type-I 2HDM with CPV. In such a model with two active doublets contributing to the EWSB, the amount of CPV is very constrained. Moreover, such a model does not provide a DM candidate. 
We further extend the scalar sector to a 3HDM with two inert and one active doublet. The active doublet plays the role of the SM-Higgs doublet. The extended inert sector accommodates DM and an unbounded amount of CPV. We have studied DM phenomenology of such a model.
\vspace{3mm}

%Table~\ref{motivation}, lists different scalar extensions of the SM and their limitations. 
%
%
%
%
%\begin{table}[t]
%\begin{center}
%\begin{tabular}{|c||c|c|}
%\hline 
%& CPV & DM \\  \hline
%2HDM  & $\surd$ & $\times$ \\ \hline
%IDM  & $\times$ & $\surd$ \\ \hline
%3HDM: I(1+2)HDM  & $\surd$ & $\surd$ \\ 
%\hline
%\end{tabular} 
%\end{center}
%\caption{Different scalar extensions of the SM and their limitations.} 
%\label{motivation}
%\end{table}

\textbf{Acknowledgements} ~The author's research is financially supported by the Academy of Finland project ``The Higgs Boson and the Cosmos'' and  project 267842. She also acknowledges the H2020-MSCA-RICE-2014 grant no. 645722 (NonMinimalHiggs).


\begin{thebibliography}{99}
\bibitem{Aad:2012tfa}
{ATLAS} Collaboration, %G.~Aad {\em et.~al.}, 
%{\it {Observation of a new
  %particle in the search for the Standard Model Higgs boson with the ATLAS
  %detector at the LHC}},  
{\em Phys.Lett.} {\bf B716} (2012) 1.%--29,
%  [\href{http://xxx.lanl.gov/abs/1207.7214}{{\tt arXiv:1207.7214}}].

\bibitem{Chatrchyan:2012ufa}
{CMS} Collaboration, %S.~Chatrchyan {\em et.~al.}, %{\it {Observation of a
  %new boson at a mass of 125 GeV with the CMS experiment at the LHC}},  
{\em
  Phys.Lett.} {\bf B716} (2012) 30.%--61,
%  [\href{http://xxx.lanl.gov/abs/1207.7235}{{\tt arXiv:1207.7235}}].



\bibitem{Jungman:1995df}
G.~Jungman, M.~Kamionkowski and K.~Griest, %{\it {Supersymmetric dark matter}},
   {\em Phys.Rept.} {\bf 267} (1996) 195.%--373,
%  [\href{http://xxx.lanl.gov/abs/hep-ph/9506380}{{\tt hep-ph/9506380}}].

\bibitem{Bertone:2004pz}
G.~Bertone, D.~Hooper and J.~Silk, %{\it {Particle dark matter: Evidence,
%  candidates and constraints}},  
{\em Phys.Rept.} {\bf 405} (2005) 279.%--390,
%  [\href{http://xxx.lanl.gov/abs/hep-ph/0404175}{{\tt hep-ph/0404175}}].

\bibitem{Bergstrom:2000pn}
L.~Bergstrom, %{\it {Nonbaryonic dark matter: Observational evidence and
%  detection methods}}, 
 {\em Rept.Prog.Phys.} {\bf 63} (2000) 793.
%  [\href{http://xxx.lanl.gov/abs/hep-ph/0002126}{{\tt hep-ph/0002126}}].


%\cite{Ade:2015xua}
\bibitem{Ade:2015xua}
  P.~A.~R.~Ade {\it et al.} [Planck Collaboration],
  %``Planck 2015 results. XIII. Cosmological parameters,''
  arXiv:1502.01589 [astro-ph.CO].
  %%CITATION = ARXIV:1502.01589;%%
  %464 citations counted in INSPIRE as of 25 Jul 2015
  

\bibitem{dimopoulos}
S.~Dimopoulos and L.~Susskind,
  %``Baryon Asymmetry in the Very Early Universe,''
  Phys.\ Lett.\ B {\bf 81}, 416 (1979)
  %doi:10.1016/0370-2693(79)90366-6
  %%CITATION = doi:10.1016/0370-2693(79)90366-6;%%
  %98 citations counted in INSPIRE as of 18 Sep 2016


%\cite{Cline:2006ts}
\bibitem{Cline:2006ts} 
  J.~M.~Cline,
  %``Baryogenesis,''
  hep-ph/0609145.
  %%CITATION = HEP-PH/0609145;%%
  %148 citations counted in INSPIRE as of 18 Sep 2016  


\bibitem{Deshpande:1977rw}
N.~G. Deshpande and E.~Ma, 
%``{Pattern of Symmetry Breaking with Two Higgs  Doublets},'' 
{\em Phys.Rev.} {\bf D18} (1978) 2574.



\bibitem{Ma:2006km}
E.~Ma,
% {\it {Verifiable radiative seesaw mechanism of neutrino mass and dark  matter}},  
{\em Phys.Rev.} {\bf D73} (2006) 077301,
  %[\href{http://xxx.lanl.gov/abs/hep-ph/0601225}{{\tt hep-ph/0601225}}].


\bibitem{Barbieri:2006dq}
R.~Barbieri, L.~J. Hall and V.~S. Rychkov, 
%``Improved naturalness with a heavy  higgs: An alternative road to lhc physics,'' 
{\em Phys. Rev.} {\bf D74} (2006) 015007. %hep-ph/0603188.
  
\bibitem{LopezHonorez:2006gr}
L.~Lopez~Honorez, E.~Nezri, J.~F. Oliver and M.~H.~G. Tytgat, 
%``The inert  doublet model: An archetype for dark matter,'' 
{\em JCAP} {\bf 0702} (2007) 028. % hep-ph/0612275.


\bibitem{four-types}
  V.~D.~Barger, J.~L.~Hewett and R.~J.~N.~Phillips,
  %``New Constraints On The Charged Higgs Sector In Two Higgs Doublet Models,''
  Phys.\ Rev.\ D {\bf 41} (1990) 3421; 
  %%CITATION = PHRVA,D41,3421;%%
%
  Y.~Grossman,
  %``Phenomenology of models with more than two Higgs doublets,''
  Nucl.\ Phys.\ B {\bf 426} (1994) 355;
  %%CITATION = HEP-PH/9401311;%%
%
  A.~G.~Akeroyd,
  %``Nonminimal neutral Higgs bosons at LEP-2,''
  Phys.\ Lett.\ B {\bf 377} (1996) 95.
%  [hep-ph/9603445].
  %%CITATION = HEP-PH/9603445;%%
  %45 citations counted in INSPIRE as of 31 Jan 2014

\bibitem{typeX} 
  M.~Aoki, S.~Kanemura, K.~Tsumura and K.~Yagyu,
  %``Models of Yukawa interaction in the two Higgs doublet model, and their collider phenomenology,''
  Phys.\ Rev.\ D {\bf 80} (2009)  015017. 
%  [arXiv:0902.4665 [hep-ph]].
  %%CITATION = ARXIV:0902.4665;%%
  %141 citations counted in INSPIRE as of 25 Feb 2015


  
\bibitem{HiggsBasis} 
  S.~Davidson and H.~E.~Haber,
  %``Basis-independent methods for the two-Higgs-doublet model,''
  Phys.\ Rev.\ D {\bf 72}, 035004 (2005)
  [Phys.\ Rev.\ D {\bf 72}, 099902 (2005)]. 
%  [hep-ph/0504050].
  %%CITATION = HEP-PH/0504050;%%
  %154 citations counted in INSPIRE as of 12 Oct 2015
  
  
   %\cite{Keus:2015hva}
\bibitem{Keus:2015hva} 
  V.~Keus, S.~F.~King, S.~Moretti and K.~Yagyu,
  %``CP Violating Two-Higgs-Doublet Model: Constraints and LHC Predictions,''
  JHEP {\bf 1604}, 048 (2016)
%  doi:10.1007/JHEP04(2016)048
  [arXiv:1510.04028 [hep-ph]].
  %%CITATION = doi:10.1007/JHEP04(2016)048;%%
  %3 citations counted in INSPIRE as of 24 Jul 2016
  
  
  \bibitem{Peskin-Takeuchi}
  M.~E.~Peskin and T.~Takeuchi,
  %``A New constraint on a strongly interacting Higgs sector,''
  Phys.\ Rev.\ Lett.\  {\bf 65}, 964 (1990); 
  %``Estimation of oblique electroweak corrections,''
  Phys.\ Rev.\  D {\bf 46}, 381 (1992). 
  %%CITATION = PHRVA,D46,381;%%


\bibitem{STU-THDM} 
%  [arXiv:hep-ph/0101342].
  %%CITATION = PHRVA,D64,093003;%%
  W.~Grimus, L.~Lavoura, O.~M.~Ogreid and P.~Osland,
  %``The Oblique parameters in multi-Higgs-doublet models,''
  Nucl.\ Phys.\ B {\bf 801}, 81 (2008);
 
  H.~E.~Haber and D.~O'Neil,
  %``Basis-independent methods for the two-Higgs-doublet model III: The CP-conserving limit, custodial symmetry, and the oblique parameters S, T, U,''
  Phys.\ Rev.\ D {\bf 83}, 055017 (2011). 
%  doi:10.1103/PhysRevD.83.055017
 % [arXiv:1011.6188 [hep-ph]].
  %%CITATION = doi:10.1103/PhysRevD.83.055017;%%
  %48 citations counted in INSPIRE as of 25 Jan 2016 
  


  
%\cite{Ivanov:2011ae}
\bibitem{Ivanov:2011ae} 
  I.~P.~Ivanov, V.~Keus and E.~Vdovin,
  %``Abelian symmetries in multi-Higgs-doublet models,''
  J.\ Phys.\ A {\bf 45}, 215201 (2012)
%  doi:10.1088/1751-8113/45/21/215201
  [arXiv:1112.1660 [math-ph]].
  %%CITATION = doi:10.1088/1751-8113/45/21/215201;%%
  %24 citations counted in INSPIRE as of 24 Jul 2016



  
  %\cite{Ivanov:2012hc}
\bibitem{Ivanov:2012hc} 
  I.~P.~Ivanov and V.~Keus,
  %``Z_p scalar dark matter from multi-Higgs-doublet models,''
  Phys.\ Rev.\ D {\bf 86}, 016004 (2012)
%  doi:10.1103/PhysRevD.86.016004
  [arXiv:1203.3426 [hep-ph]].
  %%CITATION = doi:10.1103/PhysRevD.86.016004;%%
  %17 citations counted in INSPIRE as of 25 Jul 2016



%\cite{Cordero-Cid:2016krd}
\bibitem{Cordero-Cid:2016krd} 
  A.~Cordero-Cid, J.~Hernández-Sánchez, V.~Keus, S.~F.~King, S.~Moretti, D.~Rojas and D.~Sokołowska,
  %``CP violating scalar Dark Matter,''
  JHEP {\bf 1612}, 014 (2016)
%  doi:10.1007/JHEP12(2016)014
  [arXiv:1608.01673 [hep-ph]].
  %%CITATION = doi:10.1007/JHEP12(2016)014;%%


%\cite{Keus:2014jha}
\bibitem{Keus:2014jha}
  V.~Keus, S.~F.~King, S.~Moretti and D.~Sokolowska,
  %``Dark Matter with Two Inert Doublets plus One Higgs Doublet,''
  JHEP {\bf 1411} (2014) 016.
%  [arXiv:1407.7859 [hep-ph]].
  %%CITATION = ARXIV:1407.7859;%%
  %9 citations counted in INSPIRE as of 21 Jun 2015


%\cite{Keus:2015xya}
\bibitem{Keus:2015xya} 
  V.~Keus, S.~F.~King, S.~Moretti and D.~Sokolowska,
  %``Observable Heavy Higgs Dark Matter,''
  JHEP {\bf 1511}, 003 (2015)
%  doi:10.1007/JHEP11(2015)003
  [arXiv:1507.08433 [hep-ph]].
  %%CITATION = doi:10.1007/JHEP11(2015)003;%%
  %3 citations counted in INSPIRE as of 24 Jul 2016
  
  


\end{thebibliography}
\end{document}